# Genetic Regulation of Cytokine Response in Patients with Acute Community-acquired Pneumonia

Andreas Kühnapfel[1], Katrin Horn[1], Ulrike Klotz[1], Michael Kiehntopf[2], Maciej Rosolowski[1], Markus Loeffler[1], Peter Ahnert[1], Norbert Suttorp[3], Martin Witzenrath[3], Markus Scholz[1]

[1]Institute for Medical Informatics, Statistics and Epidemiology, Medical Faculty, Leipzig University, 04103 Leipzig, Saxony, Germany

[2]Institute for Clinical Chemistry and Laboratory Diagnostics, Jena University Hospital, 07740 Jena, Thuringia, Germany

[3]Division of Infectiology and Pneumonology, Medical Department, Charité – Berlin University Medicine, 13353 Berlin, Berlin, Germany

## Abstract

Background: Community-acquired pneumonia (CAP) is an acute disease condition with a high risk of rapid deteriorations. We analysed the influence of genetics on cytokine regulation to obtain a better understanding of patient's heterogeneity.

Methods: For up to N=389 genotyped participants of the PROGRESS study of hospitalised CAP patients, we performed a genome-wide association study of ten cytokines IL-1β, IL-6, IL-8, IL-10, IL-12, MCP-1 (MCAF), MIP-1α (CCL3), VEGF, VCAM-1, and ICAM-1. Consecutive secondary analyses were performed to identify independent hits and corresponding causal variants.

Results: 102 SNPs from 14 loci showed genome-wide significant associations with five of the cytokines. The most interesting associations were found at 6p21.1 for VEGF ($p=1.58 \times 10^{-20}$), at 17q21.32 ($p=1.51 \times 10^{-9}$) and at 10p12.1 ($p=2.76 \times 10^{-9}$) for IL-1β, at 10p13 for MIP-1α (CCL3) ($p=2.28 \times 10^{-9}$), and at 9q34.12 for IL-10 ($p=4.52 \times 10^{-8}$). Functionally plausible genes could be assigned to the majority of loci including genes involved in cytokine secretion, granulocyte function, and cilial kinetics.

Conclusions: This is the first context-specific genetic association study of blood cytokine concentrations in CAP patients revealing numerous biologically plausible candidate genes. Two of the loci were also associated with atherosclerosis with probable common or consecutive pathomechanisms.

# Introduction

Community-acquired pneumonia (CAP) is an acute inflammatory condition of the lung acquired outside of the health care system. It affects people of all ages. The disease is characterised by a risk of rapid deterioration with high mortality, which is difficult to predict. Thus, hospitalisation and narrow surveillance of patients is often required (Lanks et al. 2019).

CAP has high inter-individual heterogeneity due to the complex regulation of the immune system comprising highly non-linear dynamics (Curran et al. 2014). Cytokines released during inflammatory response were shown predictive for treatment failure and mortality (Fernández-Serrano et al. 2003; Ioanas et al. 2004). We showed in the past that cytokine dynamics are causally related to relevant clinical outcome parameters (Rosolowski et al. 2020).

Genetic determinants of immune response are poorly investigated due to the fact that cross-sectional data of cytokines in population-based cohorts are less informative for acute conditions while patients with acute disease are particularly difficult to collect. Genome-wide association analyses comprised the impact of MCP-1 on the risk of stroke (Georgakis et al. 2019), the pharmacogenomics of rheumatoid arthritis treatment using anti-TNF therapy (Bek et al. 2017), the causal role of cytokines in immune-related and chronic diseases (Ahola-Olli et al. 2017), the comorbidity of schizophrenia with tuberculosis identifying common cytokines involved (Cai et al. 2016), and pleiotropic effects on cytokines (Nath et al. 2019).

We established the PROGRESS study collecting data of 3,000 hospitalised CAP patients at baseline and for four to five consecutive days (Ahnert et al. 2016). Using this resource, we aim to unravel genetic determinants of cytokine response of CAP patients. We performed a genome-wide association study, and consecutively, secondary analyses to identify novel loci of context-specific cytokine response and to corroborate other candidate loci.

# Methods

## Study Sample

Participants were recruited within the framework of the PROGRESS study (clinicaltrials.gov identifier: NCT02782013). PROGRESS is a multi-center clinical observational study of hospitalised patients with CAP. Details of the study design and inclusion/exclusion criteria can be found in (Ahnert et al. 2016). In brief, patients were included if they were 18 or more years old and had a working diagnosis of pneumonia. Patients were excluded if they stayed in hospital during the previous 28 days or if they were hospitalised for more than 48 hours before enrolment to avoid recruitment of patients with nosocomial infections, i.e. hospital-acquired pneumonia. Patients with HIV infection, AIDS, or immunosuppressive treatments within the past six months, pregnancy, or other lung diseases were also excluded. Data collection includes daily measurements of parameters of disease severity such as the Sequential Organ Failure Assessment (SOFA) score and laboratory parameters (Ahnert et al. 2019). Additionally, patients were also characterised for a number of molecular layers including genetics, transcriptomics, and proteomics.

## Cytokine Measurement and Analysis

Cytokines were measured in serum by a LUMINEX based multiplex Bead Array System (Luminex 200). We here determined the cytokines IL-1β, IL-6, IL-8, IL-10, IL-12, MCP-1 (MCAF), MIP-1α (CCL3), and VEGF (Bioplex Pro human cytokine Group I 14-Plex Kit), and VCAM-1 and ICAM-1 (Bioplex Pro human cytokine Group II 2-Plex Kit). Cytokine measurement was performed within 11 batches for a total of 403 patients. We only considered measurements at the day of inclusion of each patient. This results in a sample size of N=400. Details of the measurements can be found in (Rosolowski et al. 2020). Baseline statistics of study participants are comprised in table 1.

Table 1: Overview of patient characteristics. Statistics are presented as median (minimum to maximum), respectively absolute numbers (percentages). Total number of patients with cytokine measurement at baseline is N=400.

| Trait | Statistics |
|---|---|
| Age [years] | 62 (18-94) |
| Males / Females | 239 (60%) / 161 (40%) |
| Body mass index [kg/m$^2$] | 26.3 (15.0-54.3) |
| Current smoking | 117 (29%) |
|   Years of smoking (current and former smoker) | 14 (0-65) |
|   Years of smoking (only current smoker) | 25 (4-60) |
| Chronic kidney disease | 39 (10%) |
| Chronic liver disease | 9 (2%) |
| Diabetes | 83 (21%) |
| Antibiotic therapy prior to hospitalisation | 100 (25%) |

Data analyses was carried out cytokine-wise. Values below the limit of detection (LOD) were set to missing. Values not under LOD were transformed by natural logarithm. Outlier detection was performed by exclusion of values more than four inter-quartile ranges above the third quartile respectively below the first quartile. This removed 1 sample for VEGF and 2 samples for VCAM-1 and ICAM-1 each. We adjusted for batch effects using Empirical Bayes method as implemented in the package ComBat (Johnson et al. 2007) of the statistical software suite "R".

## Genotyping, Quality Control, and Imputation

Genotypes were measured by the CAP2 array which is a customised SNP microarray based on the Axiom platform (Affymetrix, Santa Clara, California, USA). The CAP2 array comprises the standard genome-wide content of the Axiom Genome-Wide CEU 1 Array Plate and about 60,000 custom SNPs resulting in a total of 659,675 SNPs. Custom SNPs were selected from the literature and eQTL, GWAS, and functionally relevant variant data bases.

A total of N=2,277 samples of PROGRESS were genotyped. Genotype calling was performed with Affymetrix Power Tools (APT) software (version 2.10.2.2) with standard settings. Sample and SNP quality control (QC) were performed using the software R (version 3.5.2). During sample QC, samples were excluded if one or more of the following filtering criteria were fulfilled: Dish QC (signal to noise ratio) <0.82, call rate <0.97, differences between submitted and genotyped sex, and implausible relatedness. Genetic heterogeneity was assessed by principal component analysis (PCA). Outliers were excluded if six standard deviations away from the mean. For SNP QC, we excluded SNPs with a call rate <0.97, Fisher's Linear Discriminant (FLD) <3.6, Heterozygous Cluster Strength Offset (HetSO) <-0.1, Homozygote Ratio Offset (HomRO) <-0.9 (for three clusters), violation of Hardy–Weinberg equilibrium (HWE) ($p \leq 10^{-6}$ in exact test), and plate association ($p \leq 10^{-7}$ in $\chi^2$-test). After QC, a total of N=2,174 samples and M=600,567 SNPs were available.

IMPUTE2 software (version 2.3.2) was used for imputation along with the 1000 Genomes Project reference data base (phase 3, version 5) (Auton et al. 2015). Imputation increased the number of SNPs to M=85,064,535. We only considered associations for SNPs with minor allele frequency (MAF) ≥0.01 and imputation info score ≥0.8 resulting in M=9,140,487 markers.

### Genome-wide Association Study

Combined genotype and cytokine data were available for a minimum of N=361 and a maximum of N=389 samples depending on the cytokine (supplementary table 1). Associations between genotypes and cytokines were analysed by an additive linear regression model using software PLINK (v2.00a2LM AVX2 Intel (28 Oct 2018)). X-chromosomal markers were analysed assuming total X-inactivation (i.e. male genotypes are coded as 0/2 while female genotypes are coded as 0/1/2). P-values less than or equal to $5\times10^{-8}$ were considered genome-wide significant. Suggestive SNPs are defined by p-values larger than $5\times10^{-8}$ but less than or equal to $1\times10^{-6}$. Top-hits were priority pruned by applying a linkage disequilibrium (LD) cut-off of $r^2 \geq 0.3$ using the 1000 Genomes Project reference data base (phase 3, version 5) (Auton et al. 2015) as LD reference.

SNPs were annotated by nearby genes (nearest three genes within ±250 kilo base pairs, kb) using Ensembl (Aken et al. 2017), with other trait associations by LD-based lookup ($r^2 \geq 0.3$) in the GWAS Catalog (MacArthur et al. 2017), and with expression quantitative trait loci (eQTLs) by LD-based lookup ($r^2 \geq 0.3$) using the Genotype-Tissue Expression data base (GTEx) (Human genomics. The Genotype-Tissue Expression (GTEx) pilot analysis: multitissue gene regulation in humans 2015) and (updated) own data (Kirsten et al. 2015).

### Conditional and Joint Analysis

To identify further independent hits per locus, we considered the best associated trait and performed conditional analyses by applying the tool GCTA (version 1.92.0beta3) (Yang et al. 2011). First, we performed stepwise model selection ("cojo-slct") to identify the independent variants per locus. As LD reference panel we used the complete set of genotypes (N=2,174). In case of multiple variants per locus, conditional effect estimates were calculated using "cojo-cond".

### Credible Set Analysis

After determination of the independent signals, we aimed at identifying the respective set of SNPs containing the causal variant with high certainty. For this purpose, we considered the set of SNPs within ±500 kb of the independent lead SNPs and their respective (conditional) effect estimates and standard errors (Wakefield 2007, 2009). We then calculated respective Approximate Bayes Factors (ABF) by applying the R-package "gtx". The required prior distribution of the standard deviation was constructed empirically by the difference of the 97.5[th] and the 2.5[th] percentile of SNP effects of the respective locus divided by 2x1.96. In our data, this quantity ranged in between 0.1485 (locus 2p16.3) and 0.3074 (locus 18q21.2).

### Colocalisation Analysis

We tested whether the independent loci coincide with loci of eQTLs of candidate genes in whole blood. EQTLs were retrieved from GTEx Analysis V8 (dbGaP Accession phs000424.v8.p2) (The Genotype-Tissue Expression (GTEx) project 2013). Colocalisation analysis evaluates the posterior probability of five hypotheses ($H_0$: no associations within locus; $H_{1,2}$: associations with either trait 1 (cytokine) or trait 2 (gene expression) only, $H_3$: association with both traits but different SNPs (no colocalisation), $H_4$:

association with both traits with the same SNP – evidence for colocalisation). We consider a minimum posterior probability of 0.75 as sufficient to support one of the hypotheses. Loci were again defined by a ±500kb window around the respective lead SNPs.

### Lookup of Cytokine Coding Genes

We searched for associations in the genes coding for the cytokines analysed. This is performed by considering SNPs in the respective gene body with a ±500kb margin around gene start and stop using Genome Reference Consortium Human Build 38. To account for multiple testing, we performed Benjamini-Hochberg procedure in a hierarchical manner.

## Results

### Genome-wide Association Study and Secondary Analyses

In our GWAS analysis of ten cytokines, no signs of general inflation of test statistics were detected ($\lambda$ in between 0.9934 to 1.0140, c.f. supplementary table 1). We found 102 SNPs genome-wide significantly associating with at least one cytokine. Five of the ten cytokines were involved in genome-wide associations. SNPs could be assigned to 14 genomic loci. For all loci, there was only one independent variant according to conditional and joint analyses (CoJo-Slct). Colocalisation with blood eQTLs was found for only one locus.

A visual overview of GWAS results is given by the Manhattan plot across all cytokines in figure 1 and corresponding locus-wise statistics are provided in table 2 (for a comprehensive overview of the 14 loci we provide supplementary table 2). Regional association plots of all loci are provided as supplementary figure 1.

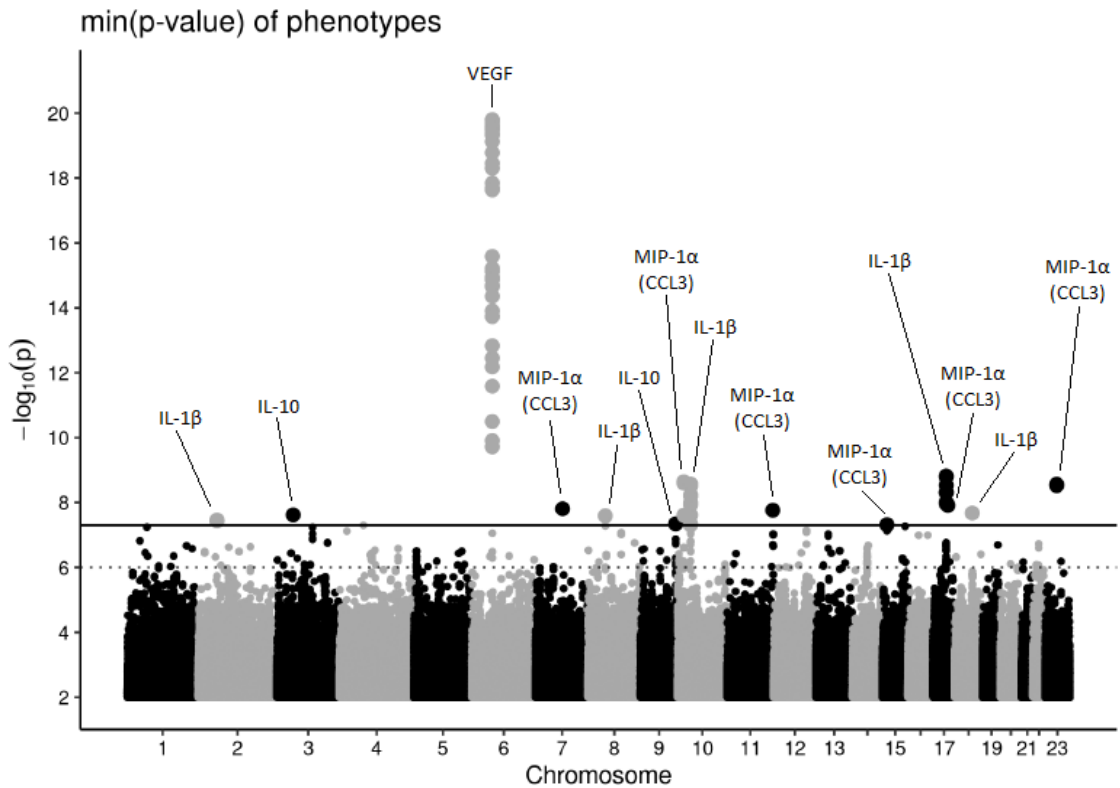

*Figure 1: Manhattan plot showing association results of the ten considered cytokines. For each SNP the maximum negative log-p-value with respect to all cytokines is shown. The solid line corresponds to the genome-wide significance threshold ($5\times10^{-8}$). The dotted line indicates suggestive associations with a p-value less than or equal to $1\times10^{-6}$ for one of the cytokines. Associations could be assigned to 14 distinct loci. Genome-wide significant hits are annotated by their best associated cytokine.*

We further illustrated the strength of association between the 14 loci (lead SNP) and all considered cytokines in figure 2. Noteworthy, only locus 6p21.1 showed genome-wide significantly association with two cytokines. For loci 3p21.31 and 9q34.12, two additional cytokines were associated with suggestive significance. All loci except for 17q22 showed nominally significant co-associations for up to six cytokines. P-value based hierarchical clustering showed grouping of cytokines across all 14 loci whereas each locus seemed to affect mainly one cytokine.

Table 2: Results of genome-wide SNP association analyses. In the table we present all 14 loci with genome-wide significant associations (threshold 5×10⁻⁸). For each locus, lead SNP, corresponding cytoband and physical position (GRCh37), physically nearby genes (within ±250 kb), best associated trait, effect allele, other allele, effect allele frequency, and beta estimate of the additive model with standard error and p-value for the top associated cytokine are shown. We also present the size of the respective 95% and 99% credible sets. For loci 3p21.31, 11q25, and Xq13.1 the lead SNPs showed minor allele frequencies below 1% in the complete data set and, thus, were excluded from the reference data set for conditional and joint analysis and credible set analysis resulting in empty cells in the corresponding rows. Loci are presented in the order of their chromosomal position.

| Locus | Lead SNP | Cytoband | Physical Position | Physical Nearby Genes (Distance to Lead SNP (kb)) | Best Associated Trait | Effect Allele | Other Allele | Allele Frequency | Beta | SE | p-value | #SNPs in credible set (first: 95%, second: 99%) |
|---|---|---|---|---|---|---|---|---|---|---|---|---|
| #1 | rs116606423 | 2p16.3 | 50123457 | RPL7P13 (17), NRXN1 (22) | IL-1β | G | T | 0.02 | 1.39 | 0.25 | 3.46 e-08 | 3,484 3,819 |
| #2 | rs139453626 | 3p21.31 | 47700006 | SMARCC1 (0), CSPG5 (78), RN7SL870P (98), DHX30 (140) | IL-10 | G | A | 0.02 | 2.42 | 0.42 | 2.39 e-08 | - |
| #3 | rs7763440 | 6p21.1 | 43926708 | C6orf223 (42), MRPL14 (150), TMEM63B (170), VEGFA (170) | VEGF | A | G | 0.45 | -0.65 | 0.07 | 1.58 e-20 | 10 12 |
| #4 | rs145122044 | 7q11.23 | 76485397 | UPK3B (0), FDPSP7 (110), DTX2P1 (120), DTX2P1-UPK3BP1-PMS2P11 (120) | MIP-1α (CCL3) | C | T | 0.01 | 1.54 | 0.27 | 1.55 e-08 | 3,230 3,631 |
| #5 | rs62505830 | 8p12 | 36121891 | RN7SKP201 (2.1), MTND6P19 (15), RNU6-533P (45) | IL-1β | T | C | 0.01 | 1.72 | 0.30 | 2.55 e-08 | 814 944 |
| #6 | rs36002018 | 9q34.12 | 133656053 | ABL1 (0), EXOSC2 (76), PRDM12 (98), QRFP (110) | IL-10 | T | A | 0.02 | 1.75 | 0.31 | 4.52 e-08 | 2,534 3,160 |
| #7 | rs75237116 | 10p13 | 12655293 | CAMK1D (0), MIR4480 (34), MIR548Q (110), | MIP-1α (CCL3) | T | C | 0.01 | 1.64 | 0.27 | 2.28 e-09 | 3,945 4,541 |

| | | | | RNU6ATAC39P (160) | | | | | | | | |
|---|---|---|---|---|---|---|---|---|---|---|---|---|
| #8 | rs6481492 | 10p12.1 | 28207393 | ARMC4 (0), RPL36AP55 (13), MPP7 (130), RN7SKP132 (130) | IL-1β | C | T | 0.06 | 0.79 | 0.13 | 2.76 e-09 | 29 50 |
| #9 | rs11223001 | 11q25 | 132161625 | NTM (0), NTM-IT (6.6), OPCML (120), RNU6-1182P (230) | MIP-1α (CCL3) | G | A | 0.01 | 1.56 | 0.27 | 1.73 e-08 | - |
| #10 | rs118008913 | 15q14 | 39036672 | C15orf53 (44), RASGRP1 (180) | MIP-1α (CCL3) | G | A | 0.02 | 1.10 | 0.20 | 4.81 e-08 | 3,397 3,761 |
| #11 | rs117439842 | 17q21.32 | 45933872 | SP6 (0.63), SCRN2 (15), LRRC46 (19), MRPL10 (25), OSBPL7 (35), SP2 (40) | IL-1β | T | G | 0.01 | 1.96 | 0.32 | 1.51 e-09 | 10 23 |
| #12 | rs8082167 | 17q22 | 52494156 | ISCA1P3 (61) | MIP-1α (CCL3) | T | C | 0.02 | 1.31 | 0.23 | 1.22 e-08 | 2,838 3,247 |
| #13 | rs76920584 | 18q21.2 | 51386971 | - | IL-1β | C | T | 0.01 | 1.49 | 0.26 | 2.11 e-08 | 194 1,687 |
| #14 | rs3788792 | Xq13.1 | 71585878 | HDAC8 (0), RNU2-68P (11), CITED1 (59), PIN4 (63) | MIP-1α (CCL3) | T | C | 0.02 | 1.17 | 0.19 | 2.74 e-09 | - |

## Known Associations

**Vascular Endothelial Growth Factor and Interleukin 12**

The strongest association was found at 6p21.1 for VEGF (rs7763440, p=1.58x10$^{-20}$). The lead SNP also showed genome-wide significance with IL-12 (strongest association for rs4320361, p=2.31x10$^{-9}$, linkage disequilibrium with lead SNP: LD=0.9976). The locus was already reported for associations with VEGF (Maffioletti et al. 2020) and blood protein levels (Sun et al. 2016). Further associations with multiple cancers (Jin et al. 2012) and ischemic stroke (especially large artery atherosclerosis, LAA) (Holliday et al. 2012) were also reported. The lead SNP is near *C6orf223*, *MRPL14*, *TMEM63B*, and *VEGFA*, where the latter is the obvious candidate gene. Furthermore, for rs7763440, we could identify the following cis-eQTL genes: *CAPN11*, *HSP90AB1*, *MRPL14*, *RSPH9*, *SLC29A1*, and *SLC35B2*. Of note, *RSPH9* was reported to be associated with primary ciliary dyskinesia (Yiallouros et al. 2019). The 99% credible set for the independent lead SNP rs7763440 comprises 12 SNPs. Another associated tag-SNP rs7739450 at this locus (p=1.92x10$^{-10}$) showed a CADD score of 10.84. The lead SNP rs7763440 colocalises with an eQTL of *C6orf223* in whole blood (posterior probability PP=93.7%).

**Interleukin-1β**

The second strongest association was rs117439842 at 17q21.32 with IL-1β (p=1.51x10$^{-9}$). The SNP is located in proximity to *SP6*, *SCRN2*, *LRRC46*, *MRPL10*, *OSBPL7*, and *SP2*. Associations with this locus were reported for this cytokine (Sherva et al. 2014) but also for primary ciliary dyskinesia (Stelzer et al. 2016) and epilepsy (genetic generalized epilepsy, genetic absence epilepsy, juvenile myoclonic epilepsy) (Steffens et al. 2012). The 99% credible set for the independent lead SNP rs117439842 consists of 23 SNPs. *SCRN2*, *LRRC46*, and *SP2* are considered as plausible candidate genes.

## Novel associations

**Interleukin-1β**

A strong association was detected at 10p12.1, again, with IL-1β (rs6481492, p=2.76x10$^{-9}$). The SNP rs6481492 is in *ARMC4* and in the near of *RPL36AP55*, *MPP7*, and *RN7SKP132*. Moreover, for the lead SNP, we could identify the cis-eQTL genes *ABI1*, *ARMC4*, *BAMBI*, *MASTL*, *RAB18*, and *WAC*. *ARMC4* was reported to be associated with primary ciliary dyskinesia (Onoufriadis et al. 2014) and vital capacity (Loth et al. 2014). For rs6481492, the 99% credible set comprises 50 SNPs. The tag-SNP rs144080867 at this locus (p=4.66x10$^{-8}$) revealed a CADD score of 11.04. In conclusion, *ARMC4* is a plausible candidate gene.

Other associations for this cytokine could be found at loci 2p16.3 (rs116606423, p=3.46x10$^{-8}$), 8p12 (rs62505830, p=2.55x10$^{-8}$), and 18q21.2 (rs76920584, p=2.11x10$^{-8}$). However, for these three loci, we cannot suggest any obvious candidate genes.

**Interleukin-10**

For IL-10, we found an association of rs36002018 at 9q34.12 (p=4.52x10$^{-8}$). The SNP is in *ABL1* and in proximity of *EXOSC2*, *PRDM12*, and *QRFP*. *ABL1* is a proto-oncogene that encodes a protein tyrosine kinase involved in a variety of cellular processes, including cell division, adhesion, differentiation, and response stress (Stelzer et al. 2016). The gene is further involved in chronic myeloid leukemia (Stelzer et al. 2016). The gene *ABL1* is a plausible candidate.

Another association was found at 3p21.31 for rs139453626 (p=2.39x10$^{-8}$). The SNP is in *SMARCC1* which belongs to the neural progenitors-specific chromatin remodeling complex (npBAF complex) and to the neuron-specific chromatin remodeling complex (nBAF complex). Nevertheless, the relationship of this gene with IL-10 needs to be elucidated.

**Macrophage Inflammatory Protein 1α**

Another association was rs75237116 at 10p13 with MIP-1α (CCL3) ($p=2.28 \times 10^{-9}$). The SNP is located in *CAMK1D* and in proximity of *MIR4480*, *MIR548Q*, and *RNU6ATAC39P*. *CAMK1D* is involved in regulation of granulocyte function, activation of *CREB* (cAMP response element binding protein)-dependent gene transcription, aldosterone synthesis, differentiation and activation of neutrophil cells, and apoptosis of erythroleukemia cells (Stelzer et al. 2016). The gene was reported to be associated with coronary artery aneurysm within Kawasaki disease (Kuo et al. 2016). We consider *CAMK1D* as the plausible candidate here.

Further associations were found at 7q11.23 (rs145122044, $p=1.55 \times 10^{-8}$), 11q25 (rs11223001, $p=1.73 \times 10^{-8}$), 15q14 (rs118008913, $p=4.81 \times 10^{-8}$), 17q22 (rs8082167, $p=1.22 \times 10^{-8}$), and at Xq13.1 (rs3788792, $p=2.74 \times 10^{-9}$). The lead SNPs on chromosomes 7, 11, and X are in the genes *UPK3B*, *NTM*, and *HDAC8*, respectively. However, biological relationships of these genes with the respective associated cytokines remain unclear.

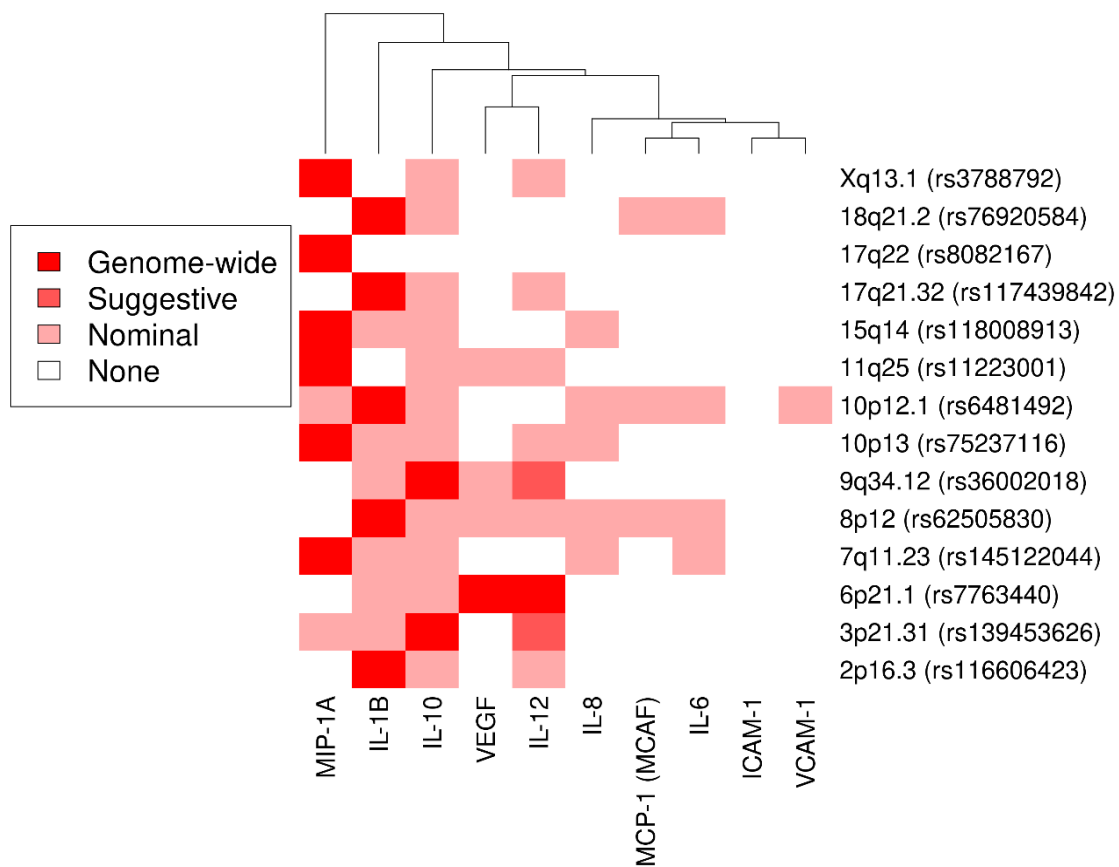

*Figure 2: Cluster heatmap for associations between the lead SNPs of the 14 loci and the 10 cytokines of interest. Strength of association is indicated by color intensity. Cytokines are ordered by hierarchical clustering according to p-value similarity.*

## Lookup of Cytokine-coding Genes

We could identify significant associations for 40 SNPs in or nearby the corresponding gene from a total of 24,354 SNPs applying hierarchical false discovery rate control at 5%. Only two cytokines were involved in these associations, i.e. associations were found for only two candidate loci. Results can be found in supplementary table 2. A total of 39 of these SNPs correspond to VEGF at locus 6p21.1. For 6p21.1, we found associations with multiple cancers (Jin et al. 2012) and large artery atherosclerotic stroke (Holliday et al. 2012) in the literature. One significant SNP corresponds to MIP-1α (CCL3) on locus 17q12. The locus reveals associations with acute lymphoblastic leukemia (Wiemels et al. 2018), cervical cancer (Shi et al. 2013), and the fraction of exhaled nitric oxide values (van der Valk et al. 2014). No associations were found for the coding genes of the other eight cytokines.

## Discussion

CAP is a disease affecting people of all ages with high mortality. Cytokines are of potential value to predict the future disease course but their context specific genetics is only partly understood. In this work, we performed a genome-wide association study and consecutive fine-mapping to elaborate genetic determinants of ten major cytokines measured in blood serum.

Among the ten cytokines there were five interleukins (1β, 6, 8, 10, and 12). Pro- and anti-inflammatory cytokines IL-6, IL-8, and IL-10 play an important role in the complex response of the human immune system due to CAP (Endeman et al. 2011). The influence of IL-12 on the expression and signaling pathways of VEGF and consequently on angiogenesis has already been demonstrated in tumour cells by Dias et al. (Dias et al. 1998) and for type 2 diabetes mellitus by Ali et al. (Ali et al. 2017). An interaction of IL-12 with IL-10 could also be demonstrated. IL-12 activates the immune response of the T-helper cells type 1, characterised by the cytokine interferon-γ, and mediates the activation of macrophages for the elimination of the pathogen. This process is negatively regulated by IL-10 (T helper cell type 2). The ratio of IL-10/IL-12 can be used as a measure for the balance of pro- and anti-inflammatory mediators and thus can be used to assess the immune status (O'Garra und Murphy 2009).

In our GWAS, we found genome-wide associations for five of the analysed ten cytokines, namely VEGF, IL-12, IL-1β, IL-10, and MIP-1α (CCL3). Genetic associations with cytokine VEGF at locus 6p21.1 have already been investigated in some studies. The two GWAS by Debette et al. (Debette et al. 2011) and Ahola-Olli et al. (Ahola-Olli et al. 2017) report a strong association of the variant rs6921438 (top SNP rs7763440; LD=0.93) with VEGF concentration. The SNP is located 171 kb downstream of the *VEGFA* gene, which codes for VEGF. Thus, *VEGFA* is the plausible candiate gene. The known variant at this locus is also believed to influence the concentration of four other cytokines (IL-12, IL-7, IL-10, and IL-13) (Ahola-Olli et al. 2017). We could confirm these results by showing genome-wide significance of this locus with VEGF and IL-12 and nominal significance for IL-10. The cytokines IL-7 and IL-13 were not considered in our study.

For the cytokine IL-1β, the variants rs117439842 and rs9903904 at locus 17q21.32 were found in the present study. The latter association also confirms results of another GWAS (Sherva et al. 2014). The following candidate genes were discerned: *SCRN2*, *LRRC46*, and *SP2*. The *SP6* gene is associated with Hermansky-Pudlak syndrome (HPS), which is associated with pulmonary fibrosis. HPS is characterized by dysregulation of alveolar macrophages (AM) (Rouhani et al. 2009). AM are known to secrete the cytokine IL-1β (Borish et al. 1992). According to Gene Ontology annotation, *OSBPL7* is involved in the binding of cholesterol, while *MRPL10* plays a role in mitochondrial translation and the translation of viral mRNA. The latter is crucial for the spread of viruses in the organism. However, the genes cannot be directly linked to CAP or IL-1β.

The gene *CAMK1D* appears in connection with two genome-wide significant SNPs at locus 10p13 for the cytokine MIP-1α (CCL3). The variant found, rs7902334, was already identified in 2016 by Kuo et al. (Kuo et al. 2016) in a GWAS on Kawasaki syndrome. *CAMK1D* is a protein-coding gene of the calcium/calmodulin-dependent protein kinases 1 family. As part of the CaMKK-CaMK1 signalling cascade, it regulates, among other things, the calcium-mediated granulocyte function and the activation of CREB-dependent transcription. In patients with CAP, a reduction in the level of CREB-regulated transcription is observed during recovery (Voevodin et al. 2019). We therefore consider *CAMK1D* as the plausible candidate.

Several genes (*ARMC4*, *SP2*, *LRRC46*, *RSPH9*, and *ZMYND10*) assigned to associations are involved in primary ciliary dyskinesia (PCD), an autosomal recessive inherited disease. Mutations of a pool of approximately 250 genes can lead to structural and/or functional dysfunction of the motile cilia. The structure of these hair-like cellular extensions of epithelial cells in the respiratory tract follows a fixed pattern of microtubules (9+2) and associated structures including dynein arms and spokes. In the respiratory tract, motile cilia are responsible for the removal of mucus, thereby protecting it from infection and ensuring mucociliary clearance. Hjeij et al. have shown that the *ARMC4* gene plays an important role in anchoring the outer dynein arms. In the case of a defect, the affected cilia exhibit reduced beating frequencies and amplitudes or become immotile (Hjeij et al. 2013). Some of the SNPs in the *ARMC4* gene region have high CADD scores, so that pathogenic effects of these variants can be assumed. The substitutions promote the development of defective proteins that disrupt the correct structure of the axoneme. For the cytokine VEGF, the study also revealed the cis-eQTL gene *RSPH9* at 6p21.1. This gene also codes for a component of the motile cilia, the so-called radial spoke head. The radial spokes support the order of the tubule pairs. Mutations of *RSPH9* can cause changes in the movement of motile cilia (Castleman et al. 2009). Of note, for a SNP at locus 9q34.12, another gene with a linkage to PCD was found. This is due to a trans-eQTL with the gene *ZMYND10*. Mutations of this gene can lead to the loss of the inner and outer dyneinarm complexes and thus to immobility of the cilia (Zariwala et al. 2013). Malfunctions in the coordinated movement of cilia can impede the removal of invading microorganisms from the respiratory tract and promote infections of the respiratory tract. Therefore, PCD patients often suffer from recurrent pneumonia. In summary, several associations point to genes involved in ciliary function providing a functional link towards liability to or severity of pneumonia, and with it, cytokine regulation.

The individual´s immune responses depend on various factors and are also affected by the pathogen which is known only for a small subset of patients of our cohort. The total sample size of this study was small, which is another limitation. Replication of our results in other studies are therefore required. Although our analyses were performed context-specific (baseline values at hospitalisation), cytokine response dynamics were not analysed.

This genome-wide association study revealed 14 loci, two of them already known. Several functional genes assigned to loci are involved in primary ciliary dyskinesia making them biologically plausible. Larger sample sizes and time series are required to further corroborate and improve our findings. Functional studies should be initiated to test our candidate genes.

## Supplementary Material

Information on the supplementary material is available under the following link: https://speicherwolke.uni-leipzig.de/index.php/s/QRX2drqH5mNJGi4.

## Acknowledgement

This work was supported by the German Research Foundation (SFB-TR84 C6 and C9, SFB 1449 B2) and by the German Federal Ministry of Education and Research (BMBF) within the framework of CAPSyS (01ZX1304A, 01ZX1304B, 01ZX1304D), CAPSyS-COVID (01ZX1604B), SYMPATH (01ZX1906A, 01ZX1906B), PROVID (01KI20160A), P4C (16GW0141), MAPVAP (16GW0247), NUM-NAPKON (01KX2021).

Data and biomaterials were made available by the PROGRESS consortium from the Prospective, longitudinal, multi-center case control study on progression of community acquired pneumonia.